\documentclass[sigconf]{acmart}
\AtBeginDocument{%
  }

\usepackage[T1]{fontenc}

\renewcommand\footnotetextcopyrightpermission[1]{} 
\acmConference{}
\acmYear{}
\acmDOI{}
\acmISBN{}
\acmBooktitle{}

\usepackage{geometry}
\usepackage{tabularx}
\usepackage{booktabs}
\usepackage{array}
\usepackage{multirow}
\usepackage{caption}
\usepackage{pdflscape}
\usepackage{pgfplots}
\usepgfplotslibrary{groupplots}
\usepackage{afterpage}
\usepackage{arydshln}
\usepackage{graphicx}
\usepackage{datatool}
\usepackage{tikz}
\pgfplotsset{compat=1.17}
\usepackage{subcaption}
\usetikzlibrary{patterns}
\usepackage[title]{appendix}

\begin{document}

 \title{FinVision: A Multi-Agent Framework for Stock Market Prediction}

\author{Sorouralsadat Fatemi}
\email{sfatem6@uic.edu}
\affiliation{%
  \institution{University of Illinois at Chicago}
  \city{Chicago}
  \state{IL}
  \country{USA}
}
\author{Yuheng Hu}
\email{yuhenghu@uic.edu}
\affiliation{%
  \institution{University of Illinois at Chicago}
  \city{Chicago}
  \state{IL}
  \country{USA}
}
\begin{abstract}
  Financial trading has been a challenging task, as it requires the integration of vast amounts of data from various modalities. Traditional deep learning and reinforcement learning methods require large training data and often involve encoding various data types into numerical formats for model input, which limits the explainability of model behavior. Recently, LLM-based agents have demonstrated remarkable advancements in handling multi-modal data, enabling them to execute complex, multi-step decision-making tasks while providing insights into their thought processes. This research introduces a multi-modal multi-agent system designed specifically for financial trading tasks. Our framework employs a team of specialized LLM-based agents, each adept at processing and interpreting various forms of financial data, such as textual news reports, candlestick charts, and trading signal charts. A key feature of our approach is the integration of a reflection module, which conducts analyses of historical trading signals and their outcomes. This reflective process is instrumental in enhancing the decision-making capabilities of the system for future trading scenarios. Furthermore, the ablation studies indicate that the visual reflection module plays a crucial role in enhancing the decision-making capabilities of our framework.

\end{abstract}




\maketitle

\section{Introduction}
The complexities and volatility of financial markets, along with multi-modal data streams, present significant challenges for tasks such as trading and market movement prediction. Effective prediction and trading systems must integrate all available information comprehensively and employ sophisticated algorithmic designs to achieve superior performance \cite{2wen2019stock, 2picasso2019technical}. To improve trading systems, the field has progressed from rule-based trading strategies to more advanced deep learning models and Reinforcement Learning (RL)-based agents \cite{2koa2023diffusion, 2zhang2022decision, 2qin2024earnhft, han2023select}. However, these models face substantial challenges, including the need for extensive training data, the oversimplification of diverse financial data types, and a lack of interpretability in their decision-making processes \cite{2yang2023fingpt}.

A key challenge in these advanced models is the effective integration of diverse financial data types without oversimplification. For instance, incorporating textual news data into deep learning and RL models presents complex challenges: reducing multifaceted content to single-variable sentiment scores fails to capture market dynamics, while effectively interpreting this information requires sophisticated financial reasoning to track evolving events and market developments over time \cite{bybee2023narrative}. Similarly, representing historical price data and technical indicators poses significant challenges due to their high dimensionality, non-linear relationships, and time-dependent nature, which can lead to information loss or misinterpretation \cite{li2023explainable, sujatha2023empirical}. Attempts to address these issues by increasing the number of variables to represent different aspects of the market often result in increased model complexity, making the internal representations and decision-making processes more intricate and harder to interpret \cite{li2023explainable, sujatha2023empirical}. 


Recent advancements in Large Language Models (LLMs) have driven their evolution into agents capable of executing complex, multi-step decision-making tasks \cite{xu2023large, zhu2022solving, gou2023critic}.  This progress has expanded the potential applications of LLMs to a diverse array of challenging domains, including mathematical reasoning, software development, and scientific research \cite{qian2024chatdev, liang2024can, du2024multi, elhenawy2024visual}. To address these complex tasks, researchers have developed a methodology that decomposes them into distinct sub-tasks \cite{du2024multi}. This approach employs multiple LLM-powered agents that collaborate, each focusing on specific aspects of the overall task, to derive comprehensive solutions. By mimicking human cognitive processes, this method enhances reasoning capabilities and problem-solving efficacy.

This collaborative approach significantly addresses a critical limitation of previous deep learning and RL models by enhancing model explainability. The transparent nature of the agents' thought processes through Chain of Thought (CoT) prompting allows for step-by-step tracking of solution derivation, providing valuable insights into their decision-making \cite{zhang2022automatic}. This explainable approach not only facilitates a deeper understanding of model operations but also enables the fine-tuning of agent prompts, framework design, and task assignment. Furthermore, the emergence of multi-modal LLMs, such as GPT-4V and the cost-effective GPT-4o, has further expanded LLM capabilities by incorporating both textual and visual data \cite{openai2023gpt4}. This integration of multimodal inputs enhances the versatility and applicability of LLM-based agents across a broader range of complex tasks \cite{elhenawy2024visual}.




These advancements unlock new possibilities for comprehensive analysis across various domains, particularly in finance. In this field, integrating diverse data types—such as textual reports, news articles, and visual data like charts—is essential for making accurate trading decisions. The evolution of LLMs and multi-agent systems holds the potential to revolutionize financial analysis by providing more sophisticated approaches to understanding market dynamics.

The application of LLMs in stock prediction has been evolving, with existing studies primarily focusing on methods such as pre-trained LLMs or instruction tuning, which require extensively annotated datasets \cite{2steinert2023linking, 2yu2023temporal, 2yang2023fingpt}. In the context of LLM-based agents, FinAgent proposed a multimodal LLM trading agent with market intelligence and reflection modules \cite{zhang2024finagent}. Our study builds upon this framework by experimenting with a significantly shorter training time of just two months, which helps to reduce API costs. Furthermore, we extend the decision-making process by requiring the model to predict the position size for trading as a percentage of the portfolio, thus inducing a more granular approach to risk management and capital allocation within our framework.

\begin{figure*}[h]
  \centering
  \includegraphics[width=\linewidth]{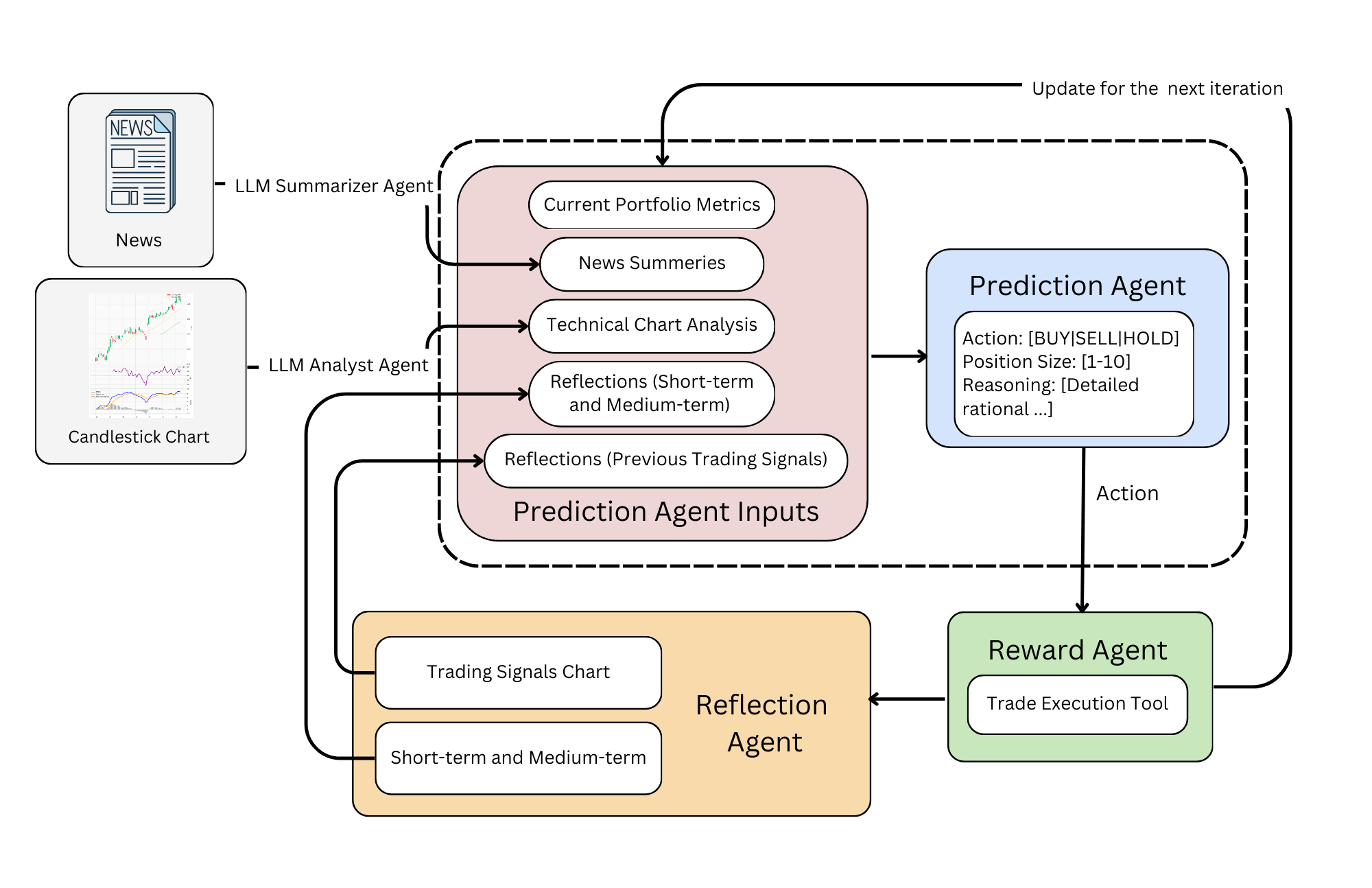}
  \caption{The Multi-modal Multi-Agent Prediction Framework}
  \label{fig1}
\end{figure*}

Our framework comprises four primary components: the Summarize Module, the Technical Analyst Module, the Prediction Module, and the Reflection Module. The Summarize Module condenses large volumes of textual news data into concise summaries that highlight factual information influencing stock trading decisions. The Technical Analyst Agent leverages the visual reasoning capabilities of LLMs to analyze candlestick charts with technical indicators, providing interpretations for next-day trading strategies. The Reflection Module consists of two parts: one assesses the short-term and medium-term performance of previous trades, while the other plots past trading signals, generates charts, and offers insights into the effectiveness of trades. The Prediction Agent integrates information from these components to forecast trading actions, determine position size as a percentage of the portfolio, and provide a detailed explanation of the decision. Based on the Prediction Agent's output, the Reward Agent executes trades and calculates performance metrics. These metrics are then used by the Reflection and Prediction Agents in the subsequent iterations. The detailed flow of our framework is illustrated in Figure \ref{fig1}.

We evaluate the performance of our framework on three major technology companies (Apple, Amazon, and Microsoft) over a seven-month period. Our findings indicate that our approach outperforms previous rule-based and reinforcement learning (RL)-based models, although it still falls short of the benchmark set by the FinAgent framework. The analysis of the trading signals reveals a comprehensive integration of diverse data sources, including financial news, candlestick chart analysis, and insights from the reflection module. This holistic approach demonstrates the potential of our framework while also highlighting areas for further development. Notably, our ablation studies underscore the significant contribution of the reflection module to overall performance.

\section{Related Works}

Large Language Models (LLMs) have demonstrated strong capabilities across various Natural Language Processing (NLP) tasks \cite{bubeck2023sparks, touvron2023llama, openai2022chatgpt, openai2023gpt4, chang2024survey}. In the finance domain, recent studies have leveraged LLMs for sentiment analysis through instruction fine-tuning, achieving superior performance compared to previous state-of-the-art models \cite{2zhang2023enhancing}. A limited number of studies have explored LLMs' potential in predicting stock market movements and conducting next-day trading through in-context learning, where historical prices and news are used to forecast future movements. However, due to the multi-modal nature of financial data and the need for multi-step reasoning, LLMs initially struggled with effective reasoning and performance in these complex tasks.

One study investigated the sentiment extracted from GPT-4 and ChatGPT and its correlation with stock price movements \cite{2steinert2023linking, 2xie2023wall}. Another study involved instruction fine-tuning LLaMA models on various financial tasks, including stock market prediction \cite{xie2024pixiu}. However, fine-tuning requires massive annotated training data, and the results often lack explainability, which makes refining and improving the models challenging.

Recently, LLM-based agents have revolutionized the field by equipping systems with advanced cognitive skills for multi-step reasoning and interaction. The use of multiple agents has been widely adopted to enhance reasoning and factuality through frameworks such as Multi-Agent Debate (MAD) and ReConcile, where multiple AI agents engage in collaborative problem-solving to improve reasoning and decision-making abilities by emulating human discussion processes \cite{du2023improving, cobbe2021training, chen2023reconcile}. In the realm of LLM-based agents, FinMem introduced an LLM trading agent with a memory mechanism that incorporates numerical historical prices but lacks agents with visual reasoning capabilities \cite{2yu2024finmem}. Another research effort proposed FinAgent, a multi-modal LLM trading agent equipped with market intelligence, low-level and high-level reflection modules, and a tool-augmented decision-making process \cite{zhang2024finagent}. While FinAgent demonstrated promising results, it required a lengthy one-year training period, leading to significant API costs. Additionally, an essential component of trading, risk management, was not accounted for in the context of this study.

Our work aims to bridge this gap by applying a multi-modal multi-agent LLM framework to the complex domain of financial trading tasks, featuring a short two-month training period and incorporating risk management into the framework.

\section{Methodology}
In this section, we outline the task definition and data specifications for our stock trading framework.

\subsection{Summary Module}
The Summary module generates concise, informative summaries from input texts. We prompt an agent to generate summaries of factual information about a specific ticker $s$ from the provided news corpora for the previous day. This process can be formalized as:
\begin{equation}
X_1^{s_{t-1}} = \text{agent}_\text{summarizer}(s, \mathcal{C}^s{t-1}, \text{prompt}_\text{summary})
\end{equation}
where $s$ is the specified stock, $\mathcal{C}^s_{t-1}$ represents the news text inputs for the previous day, $\text{agent}_\text{summarizer}$ is the language model agent generating the summary, $X_1^{s_{t-1}}$ is the generated summary, and $\text{prompt}_\text{summary}$ is the instruction given to the agent for the summarization task. This approach distills the previous day's news into concise and pertinent summaries for financial analysis. The prompt message utilized by this agent is presented in Table \ref{restPrompt} in Appendix \ref{app}.

\subsection{Technical Analyst Module}
The Technical Analyst module extracts insights from historical price data and technical indicators presented in image format. We prompt an agent to analyze the visual data and generate technical insights for a specific ticker $s$. This process can be formalized as:
\begin{equation}
X_2^{s_{t-1}} = \text{agent}_\text{technical}(s, \mathcal{I}^s{t-1}, \text{prompt}_\text{technical})
\end{equation}
where $s$ is the specified stock, $\mathcal{I}^s_{t-1}$ represents the candlestick chart and technical indicator images for the past 60 days up to day $t-1$, $\text{agent}_\text{technical}$ is the vision-capable language model agent generating the technical analysis, $X_2^{s_{t-1}}$ is the generated technical analysis, and $\text{prompt}_\text{technical}$ is the instruction given to the agent for the technical analysis task. This approach leverages the LLM's visual reasoning capabilities to interpret charts and technical indicators, identifying patterns, trends, and potential signals that could influence the stock's future performance. The Technical Analyst module complements the Summary module's textual analysis, providing a comprehensive basis for trading decisions. The prompt message utilized by this agent is presented in Table \ref{restPrompt} in Appendix \ref{app}.

\subsection{Reflection Module}
The Reflection Module consists of two parts that analyze past trading performance and signals. The first part can be formalized as:
\begin{equation}
X_3^{s_{t-1}} = \text{agent}_\text{reflection1}(s, H^s{t-L:t-1}, \text{prompt}_\text{reflection1})
\end{equation}
where $s$ is the specified stock, $H^s_{t-L:t-1}$ represents the historical trading data and performance for the past $L$ days up to day $t-1$, $\text{agent}_\text{reflection1}$ is the language model agent generating short-term and medium-term insights, $X_3^{s_{t-1}}$ is the generated reflection, and $\text{prompt}_\text{reflection1}$ is the instruction given to the agent for this task. This component provides insights into recent trading performance and decision effectiveness. The second part of the Reflection Module can be formalized as:
\begin{equation}
X_4^{s_{t-1}} = \text{agent}_\text{reflection2}(s, V^s{t-1}, \text{prompt}_\text{reflection2})
\end{equation}
where $V^s_{t-1}$ is the visual representation of trading signals for the past 30 days up to day $t-1$, generated by a plotting tool, $\text{agent}_\text{reflection2}$ is the vision-capable language model agent analyzing this visual data, $X_4^{s_{t-1}}$ is the generated feedback, and $\text{prompt}_\text{reflection2}$ is the instruction for this visual analysis task. This component offers insights into trading signal patterns and their effectiveness based on the visual data. The prompt messages utilized by these agents are presented in Table \ref{restPrompt} in Appendix \ref{app}.

\subsection{Final Decision Module}
The Final Decision module generates trading recommendations by integrating comprehensive analyses from previous modules, including news summaries, technical analysis, and reflection outcomes. The decision-making process for each stock can be formalized as:
\begin{equation}
\hat{A}^{s_t} = \text{agent}_\text{decision}(s, X_1^{s_{t-1}}, X_2^{s_{t-1}}, X_3^{s_{t-1}}, X_4^{s_{t-1}}, P^{s_{t-1}}, \text{prompt}_\text{trading})
\end{equation}
where $s$ is the specified stock, $X_1^{s_{t-1}}$ is the summary from the Summary module, $X_2^{s_{t-1}}$ is the technical analysis from the Technical Analyst module, $X_3^{s_{t-1}}$ and $X_4^{s_{t-1}}$ are the reflections from the Reflection module, $P^{s_{t-1}}$ represents the portfolio status from the previous day generated by the Reward Agent, $\text{agent}_\text{decision}$ is the language model agent specialized in decision-making, and $\text{prompt}_\text{trading}$ is the instruction for the trading decision task. The output $\hat{A}^{s_t} = (\hat{a}^{s_t}, \hat{p}^{s_t}, \hat{e}_\text{trading}^{s_t})$ consists of the recommended action $\hat{a}^{s_t} \in {\text{BUY}, \text{SELL}, \text{HOLD}}$, the position size $\hat{p}^{s_t} \in [1, 10]$ (0 if $\hat{a}^{s_t} = \text{HOLD}$), and the detailed explanation $\hat{e}_\text{trading}^{s_t}$.
This approach ensures that the trading decision benefits from the comprehensive analysis provided by all previous modules. The prompt message utilized by this agent is presented in Table \ref{predictiom} in Appendix \ref{app}.

\subsection{Implementation Details}
Our multi-agent system utilizes the LangGraph library\footnote{LangGraph: A library for building and managing multi-agent systems. Available at \url{https://www.langchain.com/langgraph}.} to implement a directed graph structure, where each node corresponds to a specialized agent. The StateGraph class is employed to define the dependencies among agents and manage the flow of information. All agents, except for the final decision agent, utilize the GPT-4o-mini model, a capable multi-modal language model, with a temperature setting of 0.3 to achieve uniform outputs. Notably, the Chart Agent and a portion of the Reflection Agent leverage the model's vision capabilities to analyze candlestick charts, technical indicators, and trading signal images. The Prediction Agent, tasked with making the final trading decision, operates using the o1-mini model, a new GPT model designed for advanced reasoning tasks, with a temperature setting of 1 (the only available option for this model). Additionally, a custom AgentState class manages the trading system's state, encapsulating all relevant trading information. This modular design facilitates flexible agent tuning or replacement while maintaining consistent multi-modal processing throughout the pipeline.

\section{Experiments}
To validate the effectiveness of our proposed multi-agent framework, we conduct comprehensive experiments comparing it against baseline models. 


\subsection{Data Collection}
Our study examines three major technology stocks—Apple (AAPL), Amazon (AMZN), and Microsoft (MSFT)—over a nine-month period from April 1, 2023, to December 29, 2023. We structured this time frame into a two-month training period (April 1 to May 31, 2023) and a seven-month testing period (June 1 to December 29, 2023). The dataset comprises news articles sourced from Yahoo Finance,\footnote{Data retrieved using eodhd.com/api/news} daily candlestick charts, technical indicators, and reflection data. The candlestick charts, technical indicators, and trading signal images for reflection were all plotted using Matplotlib and various finance libraries. Specifically, we incorporated the following technical indicators: Simple Moving Averages (10 and 50-day), Relative Strength Index (14-day period), Bollinger Bands (20-day period with 2 standard deviations), trading volume, and Moving Average Convergence Divergence (MACD). Reflection data includes trading signal images (which contain signals from previous days) and performance data from past trading activities. This initial training period was crucial for generating sufficient reflection data, ensuring that our multi-agent system had robust historical inputs for the subsequent testing phase. Table \ref{tab:dataset-statistics} presents a summary of our dataset statistics, detailing the number of trading days, news articles, charts, and technical indicators for each asset throughout the study period.

\begin{table}[h]
\centering

\begin{tabular}{lccccc}
\hline
Ticker & Period & Trading Days & News Articles \\
\hline
AAPL & Training (Apr 1 - May 31) & 42 & 1,081 \\
     & Testing (Jun 1 - Dec 29)  & 145 & 4,886 \\
AMZN & Training (Apr 1 - May 31) & 42 & 1,113  \\
     & Testing (Jun 1 - Dec 29)  & 145 & 5,556  \\
MSFT & Training (Apr 1 - May 31) & 42 & 1,897 \\
     & Testing (Jun 1 - Dec 29)  & 145 & 1,249  \\
\hline
\end{tabular}
\caption{Dataset statistics}
\label{tab:dataset-statistics}
\end{table}

\subsection{Evaluation Metrics}
To comprehensively assess the performance of our multi-agent trading system, we employ the following key metrics:
\begin{itemize}
\item \textbf{Annual Rate of Return (ARR)}: This metric provides an annualized measure of portfolio growth, calculated as:
\begin{equation}
ARR = \frac{P_T - P_0}{P_0} \times \frac{C}{T}
\end{equation}
where $T$ is the total number of trading days, $C$ is the number of trading days within a year, and $P_T$ and $P_0$ represent the final and initial portfolio values, respectively.
\item \textbf{Sharpe Ratio (SR)}: This measures risk-adjusted returns of portfolios, defined as:
\begin{equation}
\text{Sharpe Ratio} = \frac{R_p - R_f}{\sigma_p}
\end{equation}
where $R_p$ is the portfolio's average return, $R_f$ is the risk-free rate, and $\sigma_p$ is the portfolio's volatility. A higher Sharpe Ratio suggests better risk-adjusted performance.
\item \textbf{Maximum Drawdown (MDD)}: This metric measures the largest percentage decline from a historical peak in portfolio value. It is defined as:

\begin{equation}
MDD = \max_{t \in (0,T)} \left( \frac{PV_{peak,t} - PV_{t}}{PV_{peak,t}} \right)
\end{equation}

where \( PV_t = \prod_{i=1}^t \frac{V_i}{V_{i-1}} \) is the cumulative return up to time \( t \), and \( PV_{peak,t} = \max_{i \in (1,t)} PV_i \) is the highest cumulative return up to time \( t \). Here, \( V_i \) denotes the portfolio value at time \( i \).

\end{itemize}

\subsection{Benchmark Models}
To evaluate our multi-agent trading framework, we compare its performance against traditional trading strategies and advanced algorithmic approaches:
\subsubsection{Traditional Strategies}
We implement three widely-used trading strategies:
(1) Buy-and-Hold (B\&H), a passive long-term investment approach;
(2) Moving Average Convergence Divergence (MACD), utilizing trend-following momentum indicators; and
(3) KDJ with RSI Filter, combining oscillators for refined signal generation.
\subsubsection{Reinforcement Learning Models}
We employ two reinforcement learning algorithms:
(1) Proximal Policy Optimization (PPO), which optimizes trading policies while ensuring stable learning through constrained updates, and
(2) Deep Q-Network (DQN), which learns optimal action-value functions using deep neural networks to handle complex market states \cite{schulman2017proximal, mnih2013playing}.
\subsubsection{LLM-based Benchmark}
We also compare against FinAgent, a multi-modal foundational agent with tool augmentation, trained using a reinforcement learning framework on a one-year training dataset \cite{zhang2024finagent}.

\begin{table*}
\begin{center}
\begin{tabular}{ l c c c } 
\hline
\textbf{Model} & 
\begin{tabular}{ c } 
\textbf{AAPL} \\
\hline
\begin{tabular}{ c c c } 
ARR\% & SR\% & MDD\%
\end{tabular}
\end{tabular}
&
\begin{tabular}{ c } 
\textbf{MSFT} \\
\hline
\begin{tabular}{ c c c } 
ARR\% & SR\% & MDD\%
\end{tabular} 
\end{tabular}
&
\begin{tabular}{ c } 
\textbf{AMZN} \\
\hline
\begin{tabular}{ c c c } 
ARR\% & SR\% & MDD\%
\end{tabular} 
\end{tabular}
\\
\hline 
\begin{tabular}{ l } 
\textbf{Market} \\  \hline
Buy and Hold
\end{tabular}
&
\begin{tabular}{ c } 
\begin{tabular}{ c } 
\\ 
\begin{tabular}{ c c c } 
13.56 & 0.67 & 14.93
\end{tabular} 
\end{tabular}
\end{tabular}
& 
\begin{tabular}{ c } 
\\ 
\begin{tabular}{ c c c } 
22.27 & 1.01 & 12.95
\end{tabular} 
\end{tabular}
& 
\begin{tabular}{ c } 
\\ 
\begin{tabular}{ c c c } 
\underline{43.57} & 1.37 & 17.45
\end{tabular}  
\end{tabular}
\\ \hline
\begin{tabular}{ l } 
\textbf{Rule-based Methods} \\ \hline
MACD \\ RSI
\end{tabular}
& 
\begin{tabular}{ c } 
\\ 
\begin{tabular}{ c c c } 
1.47 & -0.26 & 1.33
\end{tabular}  
\\ 
\begin{tabular}{ c c c } 
4.20 & \underline{1.22} & \textbf{0.62}
\end{tabular} 
\end{tabular}
& 
\begin{tabular}{ c } 
\\ 
\begin{tabular}{ c c c } 
0.36 & -0.71 & \underline{1.67}
\end{tabular} 
\\ 
\begin{tabular}{ c c c } 
1.54 & -0.33 & \textbf{0.63}
\end{tabular} 
\end{tabular}
& 
\begin{tabular}{ c } 
\\ 
\begin{tabular}{ c c c } 
-6.40 & -1.94 & \underline{4.56}
\end{tabular} 
\\ 
\begin{tabular}{ c c c } 
2.35 &  0.20 &  \textbf{0.32}
\end{tabular} 
\end{tabular}
\\ \hline
\begin{tabular}{ l } 
\textbf{RL-based Methods} \\ \hline
PPO \\ DQN
\end{tabular}
& 
\begin{tabular}{ c } 
\\ 
\begin{tabular}{ c c c } 
7.26 & -0.42 & 7.90
\end{tabular}  
\\ 
\begin{tabular}{ c c c } 
1.22 & -0.90 & \underline{5.87}
\end{tabular} 
\end{tabular}
& 
\begin{tabular}{ c } 
\\ 
\begin{tabular}{ c c c } 
6.23 & -0.73 & 11.26
\end{tabular} 
\\ 
\begin{tabular}{ c c c } 
17.75 & -0.26 & 12.85
\end{tabular} 
\end{tabular}
& 
\begin{tabular}{ c } 
\\ 
\begin{tabular}{ c c c } 
17.15 & -0.59 & 15.39
\end{tabular} 
\\ 
\begin{tabular}{ c c c } 
22.07 & -0.46 & 19.57
\end{tabular} 
\end{tabular}
\\ \hline
\begin{tabular}{ l } 
\textbf{LLM-based Models} \\ \hline
FinAgent
\end{tabular}
& 
\begin{tabular}{ c } 
\\ 
\begin{tabular}{ c c c } 
\textbf{31.89} & \textbf{1.43} & 10.40
\end{tabular} 
\\ 
\end{tabular}
& 
\begin{tabular}{ c } 
\\ 
\begin{tabular}{ c c c } 
\textbf{44.74} & \textbf{1.79} & 5.57
\end{tabular} 
\end{tabular}
& 
\begin{tabular}{ c } 
\\ 
\begin{tabular}{ c c c } 
\textbf{65.10} & \underline{1.61} & 13.20
\end{tabular} 
\end{tabular}
\\ \hline
\begin{tabular}{ l } 
\textbf{Our Method} \\ \hline
FinVision \\ FinVision-w/o Reflection
\end{tabular}
& 
\begin{tabular}{ c } 
\\ 
\begin{tabular}{ c c c } 
\underline{14.79} & 1.20 & 14.38
\end{tabular}  
\\ 
\begin{tabular}{ c c c } 
8.84 & 0.62 & 13.42
\end{tabular} 
\end{tabular}
& 
\begin{tabular}{ c } 
\\ 
\begin{tabular}{ c c c } 
\underline{25.57} & \underline{1.41} & 13.28
\end{tabular} 
\\ 
\begin{tabular}{ c c c } 
16.99 & 1.04 & 11.79
\end{tabular} 
\end{tabular}
& 
\begin{tabular}{ c } 
\\ 
\begin{tabular}{ c c c } 
42.14 & \textbf{1.72} & 12.09
\end{tabular} 
\\ 
\begin{tabular}{ c c c } 
37.64 & 1.68 & 10.64
\end{tabular} 
\end{tabular}
\\ \hline

\end{tabular}
\end{center}
    \caption{Performance Results of All Models. The best performance metrics are indicated in bold, while the second-best metrics are underlined. Results for FinAgent are retrieved from the respective publication \cite{2yang2023fingpt}.}
\label{performance}
\end{table*}


\begin{table*}[t]
\centering
\begin{tabular}{|p{\textwidth}|}
\hline

\textbf{Recommendation:} SELL  \\
\textbf{Position Size:} 3\% of portfolio  

\textbf{Explanation:}

\textbf{Rationale for Selling:}

1. \textbf{Unrealized Profit Above Threshold:}
   \begin{itemize}
   \item \textbf{Unrealized Profit:} The current portfolio shows an unrealized profit of \$344.10, which is a \textbf{5.63\%} gain above the average ...
   \end{itemize}

2. \textbf{Technical Indicators Suggest Mixed Signals:}
   \begin{itemize}
      \item \textbf{MACD Uptrend:} While the MACD indicates an upward trend with a \textcolor{blue}{bullish} crossover signal imminent, suggesting potential ...
   \item \textbf{RSI Overbought:} The Relative Strength Index (RSI) highlights an overbought condition, which often precedes a price correction ...
   \item \textbf{KDJ Sell Signal:} The KDJ indicator is signaling a potential sell, reinforcing the cautionary stance against holding the entire position.
   \item \textbf{Bollinger Bands:} Although the price approaching the lower Bollinger Band suggests a \textcolor{blue}{buy signal} for mean reversion, the ...
   \end{itemize}

3. \textbf{Negative Market Sentiment Due to News:}
   \begin{itemize}
   \item \textbf{Patent Dispute Impact:} The recent news about Apple halting sales of the Apple Watch Series 9 and Ultra 2 in the U.S. ...
   \item \textbf{Bearish Sentiment:} The overall market sentiment is currently bearish in the short term, further supported by the stock's recent dip ...
   \end{itemize}

4. \textbf{Historical Trading Performance Insights:}
   \begin{itemize}
   \item \textbf{Missed Opportunities \& Sell Timing:} Past trading data indicates that several sell signals may have been timed too early, resulting in missed opportunities for additional gains. However, \textcolor{blue}{given the current mixed signals from technical indicators and negative news, selling a partial position helps mitigate the risk of potential short-term downturns} while retaining a portion of the investment for long-term growth prospects.
   \end{itemize}

\textbf{Strategic Balance:}

- \textbf{Profit Lock-In:} Selling \textbf{3\%} of the portfolio ensures that a portion of the gains is realized, reducing exposure to potential short-term risks without fully divesting from the position.
  
- \textbf{Maintaining Exposure:} Retaining the majority of the portfolio allows participation in any future upside, especially if the MACD uptrend materializes despite current bearish indicators.

- \textbf{Cash Reserve Compliance:} The current cash reserve stands at \textbf{39.45\%}, well above the minimum \textbf{10\%} requirement, providing ample liquidity for any opportunistic buying if favorable conditions arise.

\textbf{Conclusion:} Selling \textbf{3\%} of the Apple Inc. stock portfolio is a balanced approach to secure existing profits amidst mixed technical signals and negative market sentiment arising from recent news. This strategy aligns with the provided \textcolor{blue}{trading rules, ensuring risk management while allowing for continued participation in potential upward movements}.
\\
\hline
\end{tabular}
\caption{Sample Response from the Prediction Agent for Apple Stock}
\label{response}
\end{table*}

 \section{Main Results}
Our framework demonstrated strong performance across three major technology stocks (AAPL, MSFT, and AMZN), as shown in Table \ref{performance}. These results highlight its versatility and effectiveness as a trading strategy, especially in the context of a strongly bullish market during the test period.\\

\textbf{Comparative Performance:}
Our FinVision framework outperformed the market buy-and-hold strategy for AAPL and MSFT in terms of Annual Return Rate (ARR) and risk-adjusted returns (Sharpe Ratio). For AAPL, the framework achieved a 14.79\% ARR and a Sharpe Ratio of 1.20, compared to the market's 13.56\% ARR and 0.67 Sharpe Ratio. Similarly, for MSFT, our framework's ARR of 25.57\% and Sharpe Ratio of 1.41 surpassed the market's 22.27\% ARR and 1.01 Sharpe Ratio. Although the framework's 42.14\% ARR for AMZN slightly lagged behind the market's 43.57\% ARR, it significantly improved risk-adjusted performance, achieving a Sharpe Ratio of 1.72 (compared to the market's 1.37) and a lower Maximum Drawdown of 12.09\% (versus 17.45\% for the market). These results demonstrate the capability of our system to generate competitive returns while effectively managing risk compared to passive strategies.\\
\textbf{Performance in Bullish Markets:}
The effectiveness of the buy-and-hold strategy, particularly for AMZN (43.57\% ARR), reflects the strong upward trend of these tech stocks during the test period. This bullish environment inherently favors passive strategies, making our framework's outperformance, or near-equivalent returns to buy-and-hold, particularly noteworthy. The framework's capacity to enhance risk-adjusted metrics while maintaining competitive returns demonstrates its effectiveness in risk management, even in strongly trending markets. These results indicate that the model provides value through both return optimization and more robust risk control approaches.\\
\textbf{Superiority over RL-based Models:}
The framework demonstrated substantially higher performance compared to reinforcement learning (RL) based models, including PPO and DQN, across all evaluated stocks. For instance, with AAPL, our framework's 14.79\% ARR and 1.20 Sharpe Ratio far exceeded those of PPO (7.26\% ARR, -0.42 Sharpe Ratio) and DQN (1.22\% ARR, -0.90 Sharpe Ratio). The consistent positive Sharpe Ratios achieved by the framework, in contrast to the negative Sharpe Ratios of RL models, indicate superior risk-adjusted performance. These performance differentials suggest that the integrated approach more effectively captures complex market dynamics compared to RL approaches, particularly in trending markets.

However, our method underperformed compared to the FinAgent model, which underscores the efficiency of their extensive training data. Nevertheless, with much less training time, our framework performs well, suggesting potential for further tuning to enhance performance.\\
\textbf{Impact of Reflection Mechanism:}
The ablation study demonstrated that the reflection component significantly contributes to the framework's performance. A comparison between the full framework and the version without reflection shows substantial improvements in performance metrics across all stocks, validating the effectiveness of this adaptive learning mechanism. This component enables the framework to calibrate its strategy based on historical performance and market conditions, resulting in consistent performance across diverse stocks and market scenarios.

Our multi-agent framework provides detailed visibility into its trading decision process. As illustrated in Table \ref{response}, the example from December 19, 2023, shows the framework's ability to effectively integrate diverse information sources. When technical indicators suggested a bullish trend for Apple stock, the prediction agent incorporated news signals and reflections from previous trading signals to arrive at a more nuanced decision, which proved accurate as the stock price peaked on that day compared to the following trading day. This explainable approach not only offers insights into the decision-making process but also facilitates the optimization of the framework for enhanced performance. Furthermore, the framework's ability to suggest position sizes adds an extra layer of risk management to the trading strategy. By providing clear reasoning for its recommendations, including the specific percentage of the portfolio to trade, the framework allows for more precise control over risk exposure. This combination of explainable decision-making and dynamic position sizing underscores the framework's effectiveness, demonstrating its potential for adaptable and risk-aware trading strategies in complex market conditions

\section{Conclusions and Future Work}

In this study, we proposed a multi-modal multi-agent framework for financial trading tasks that demonstrates superior performance compared to traditional rule-based and reinforcement learning models. The framework implements a risk-controlled investment approach, achieving competitive returns while maintaining effective risk management. The reflection component emerges as a key contributor to the framework's performance, enabling adaptive learning based on historical outcomes and market conditions.
Future research will focus on extending the framework by incorporating reinforcement learning techniques, particularly through fine-tuning the policy using verbal prompts in a dynamic refinement process, as proposed by \cite{zhang2024agentpro}. This enhancement aims to improve the framework's adaptability to rapidly changing market conditions.

\bibliographystyle{ACM-Reference-Format}
\bibliography{sample-base}


\begin{thebibliography}{36}


\ifx \showCODEN    \undefined \def \showCODEN     #1{\unskip}     \fi
\ifx \showDOI      \undefined \def \showDOI       #1{#1}\fi
\ifx \showISBNx    \undefined \def \showISBNx     #1{\unskip}     \fi
\ifx \showISBNxiii \undefined \def \showISBNxiii  #1{\unskip}     \fi
\ifx \showISSN     \undefined \def \showISSN      #1{\unskip}     \fi
\ifx \showLCCN     \undefined \def \showLCCN      #1{\unskip}     \fi
\ifx \shownote     \undefined \def \shownote      #1{#1}          \fi
\ifx \showarticletitle \undefined \def \showarticletitle #1{#1}   \fi
\ifx \showURL      \undefined \def \showURL       {\relax}        \fi
\providecommand\bibfield[2]{#2}
\providecommand\bibinfo[2]{#2}
\providecommand\natexlab[1]{#1}
\providecommand\showeprint[2][]{arXiv:#2}

\bibitem[Bubeck et~al\mbox{.}(2023)]%
        {bubeck2023sparks}
\bibfield{author}{\bibinfo{person}{S{\'e}bastien Bubeck}, \bibinfo{person}{Varun Chandrasekaran}, \bibinfo{person}{Ronen Eldan}, \bibinfo{person}{Johannes Gehrke}, \bibinfo{person}{Eric Horvitz}, \bibinfo{person}{Ece Kamar}, \bibinfo{person}{Peter Lee}, \bibinfo{person}{Yin~Tat Lee}, \bibinfo{person}{Yuanzhi Li}, \bibinfo{person}{Scott Lundberg}, {et~al\mbox{.}}} \bibinfo{year}{2023}\natexlab{}.
\newblock \showarticletitle{Sparks of artificial general intelligence: Early experiments with gpt-4}.
\newblock \bibinfo{journal}{\emph{arXiv preprint arXiv:2303.12712}} (\bibinfo{year}{2023}).
\newblock


\bibitem[Bybee et~al\mbox{.}(2023)]%
        {bybee2023narrative}
\bibfield{author}{\bibinfo{person}{Leland Bybee}, \bibinfo{person}{Bryan Kelly}, {and} \bibinfo{person}{Yinan Su}.} \bibinfo{year}{2023}\natexlab{}.
\newblock \showarticletitle{Narrative asset pricing: Interpretable systematic risk factors from news text}.
\newblock \bibinfo{journal}{\emph{The Review of Financial Studies}} \bibinfo{volume}{36}, \bibinfo{number}{12} (\bibinfo{year}{2023}), \bibinfo{pages}{4759--4787}.
\newblock


\bibitem[Chang et~al\mbox{.}(2024)]%
        {chang2024survey}
\bibfield{author}{\bibinfo{person}{Yupeng Chang}, \bibinfo{person}{Xu Wang}, \bibinfo{person}{Jindong Wang}, \bibinfo{person}{Yuan Wu}, \bibinfo{person}{Linyi Yang}, \bibinfo{person}{Kaijie Zhu}, \bibinfo{person}{Hao Chen}, \bibinfo{person}{Xiaoyuan Yi}, \bibinfo{person}{Cunxiang Wang}, \bibinfo{person}{Yidong Wang}, {et~al\mbox{.}}} \bibinfo{year}{2024}\natexlab{}.
\newblock \showarticletitle{A survey on evaluation of large language models}.
\newblock \bibinfo{journal}{\emph{ACM Transactions on Intelligent Systems and Technology}} \bibinfo{volume}{15}, \bibinfo{number}{3} (\bibinfo{year}{2024}), \bibinfo{pages}{1--45}.
\newblock


\bibitem[Chen et~al\mbox{.}(2023)]%
        {chen2023reconcile}
\bibfield{author}{\bibinfo{person}{Justin Chih-Yao Chen}, \bibinfo{person}{Swarnadeep Saha}, {and} \bibinfo{person}{Mohit Bansal}.} \bibinfo{year}{2023}\natexlab{}.
\newblock \showarticletitle{Reconcile: Round-table conference improves reasoning via consensus among diverse llms}.
\newblock \bibinfo{journal}{\emph{arXiv preprint arXiv:2309.13007}} (\bibinfo{year}{2023}).
\newblock


\bibitem[Cobbe et~al\mbox{.}(2021)]%
        {cobbe2021training}
\bibfield{author}{\bibinfo{person}{Karl Cobbe}, \bibinfo{person}{Vineet Kosaraju}, \bibinfo{person}{Mohammad Bavarian}, \bibinfo{person}{Mark Chen}, \bibinfo{person}{Heewoo Jun}, \bibinfo{person}{Lukasz Kaiser}, \bibinfo{person}{Matthias Plappert}, \bibinfo{person}{Jerry Tworek}, \bibinfo{person}{Jacob Hilton}, \bibinfo{person}{Reiichiro Nakano}, {et~al\mbox{.}}} \bibinfo{year}{2021}\natexlab{}.
\newblock \showarticletitle{Training verifiers to solve math word problems}.
\newblock \bibinfo{journal}{\emph{arXiv preprint arXiv:2110.14168}} (\bibinfo{year}{2021}).
\newblock


\bibitem[Du et~al\mbox{.}(2023)]%
        {du2023improving}
\bibfield{author}{\bibinfo{person}{Yilun Du}, \bibinfo{person}{Shuang Li}, \bibinfo{person}{Antonio Torralba}, \bibinfo{person}{Joshua~B Tenenbaum}, {and} \bibinfo{person}{Igor Mordatch}.} \bibinfo{year}{2023}\natexlab{}.
\newblock \showarticletitle{Improving factuality and reasoning in language models through multiagent debate}.
\newblock \bibinfo{journal}{\emph{arXiv preprint arXiv:2305.14325}} (\bibinfo{year}{2023}).
\newblock


\bibitem[Du et~al\mbox{.}(2024)]%
        {du2024multi}
\bibfield{author}{\bibinfo{person}{Zhuoyun Du}, \bibinfo{person}{Chen Qian}, \bibinfo{person}{Wei Liu}, \bibinfo{person}{Zihao Xie}, \bibinfo{person}{Yifei Wang}, \bibinfo{person}{Yufan Dang}, \bibinfo{person}{Weize Chen}, {and} \bibinfo{person}{Cheng Yang}.} \bibinfo{year}{2024}\natexlab{}.
\newblock \showarticletitle{Multi-Agent Software Development through Cross-Team Collaboration}.
\newblock \bibinfo{journal}{\emph{arXiv preprint arXiv:2406.08979}} (\bibinfo{year}{2024}).
\newblock


\bibitem[Elhenawy et~al\mbox{.}(2024)]%
        {elhenawy2024visual}
\bibfield{author}{\bibinfo{person}{Mohammed Elhenawy}, \bibinfo{person}{Ahmad Abutahoun}, \bibinfo{person}{Taqwa~I Alhadidi}, \bibinfo{person}{Ahmed Jaber}, \bibinfo{person}{Huthaifa~I Ashqar}, \bibinfo{person}{Shadi Jaradat}, \bibinfo{person}{Ahmed Abdelhay}, \bibinfo{person}{Sebastien Glaser}, {and} \bibinfo{person}{Andry Rakotonirainy}.} \bibinfo{year}{2024}\natexlab{}.
\newblock \showarticletitle{Visual Reasoning and Multi-Agent Approach in Multimodal Large Language Models (MLLMs): Solving TSP and mTSP Combinatorial Challenges}.
\newblock \bibinfo{journal}{\emph{arXiv preprint arXiv:2407.00092}} (\bibinfo{year}{2024}).
\newblock


\bibitem[Gou et~al\mbox{.}(2023)]%
        {gou2023critic}
\bibfield{author}{\bibinfo{person}{Zhibin Gou}, \bibinfo{person}{Zhihong Shao}, \bibinfo{person}{Yeyun Gong}, \bibinfo{person}{Yelong Shen}, \bibinfo{person}{Yujiu Yang}, \bibinfo{person}{Nan Duan}, {and} \bibinfo{person}{Weizhu Chen}.} \bibinfo{year}{2023}\natexlab{}.
\newblock \showarticletitle{Critic: Large language models can self-correct with tool-interactive critiquing}.
\newblock \bibinfo{journal}{\emph{arXiv preprint arXiv:2305.11738}} (\bibinfo{year}{2023}).
\newblock


\bibitem[Han et~al\mbox{.}(2023)]%
        {han2023select}
\bibfield{author}{\bibinfo{person}{Weiguang Han}, \bibinfo{person}{Boyi Zhang}, \bibinfo{person}{Qianqian Xie}, \bibinfo{person}{Min Peng}, \bibinfo{person}{Yanzhao Lai}, {and} \bibinfo{person}{Jimin Huang}.} \bibinfo{year}{2023}\natexlab{}.
\newblock \showarticletitle{Select and trade: Towards unified pair trading with hierarchical reinforcement learning}. In \bibinfo{booktitle}{\emph{Proceedings of the 29th ACM SIGKDD Conference on Knowledge Discovery and Data Mining}}. \bibinfo{pages}{4123--4134}.
\newblock


\bibitem[Koa et~al\mbox{.}(2023)]%
        {2koa2023diffusion}
\bibfield{author}{\bibinfo{person}{Kelvin~JL Koa}, \bibinfo{person}{Yunshan Ma}, \bibinfo{person}{Ritchie Ng}, {and} \bibinfo{person}{Tat-Seng Chua}.} \bibinfo{year}{2023}\natexlab{}.
\newblock \showarticletitle{Diffusion variational autoencoder for tackling stochasticity in multi-step regression stock price prediction}. In \bibinfo{booktitle}{\emph{Proceedings of the 32nd ACM International Conference on Information and Knowledge Management}}. \bibinfo{pages}{1087--1096}.
\newblock


\bibitem[Li et~al\mbox{.}(2023)]%
        {li2023explainable}
\bibfield{author}{\bibinfo{person}{Xun Li}, \bibinfo{person}{Dongsheng Chen}, \bibinfo{person}{Weipan Xu}, \bibinfo{person}{Haohui Chen}, \bibinfo{person}{Junjun Li}, {and} \bibinfo{person}{Fan Mo}.} \bibinfo{year}{2023}\natexlab{}.
\newblock \showarticletitle{Explainable dimensionality reduction (XDR) to unbox AI ‘black box’models: A study of AI perspectives on the ethnic styles of village dwellings}.
\newblock \bibinfo{journal}{\emph{Humanities and Social Sciences Communications}} \bibinfo{volume}{10}, \bibinfo{number}{1} (\bibinfo{year}{2023}), \bibinfo{pages}{1--13}.
\newblock


\bibitem[Liang et~al\mbox{.}(2024)]%
        {liang2024can}
\bibfield{author}{\bibinfo{person}{Weixin Liang}, \bibinfo{person}{Yuhui Zhang}, \bibinfo{person}{Hancheng Cao}, \bibinfo{person}{Binglu Wang}, \bibinfo{person}{Daisy~Yi Ding}, \bibinfo{person}{Xinyu Yang}, \bibinfo{person}{Kailas Vodrahalli}, \bibinfo{person}{Siyu He}, \bibinfo{person}{Daniel~Scott Smith}, \bibinfo{person}{Yian Yin}, {et~al\mbox{.}}} \bibinfo{year}{2024}\natexlab{}.
\newblock \showarticletitle{Can large language models provide useful feedback on research papers? A large-scale empirical analysis}.
\newblock \bibinfo{journal}{\emph{NEJM AI}} \bibinfo{volume}{1}, \bibinfo{number}{8} (\bibinfo{year}{2024}), \bibinfo{pages}{AIoa2400196}.
\newblock


\bibitem[Mnih(2013)]%
        {mnih2013playing}
\bibfield{author}{\bibinfo{person}{Volodymyr Mnih}.} \bibinfo{year}{2013}\natexlab{}.
\newblock \showarticletitle{Playing atari with deep reinforcement learning}.
\newblock \bibinfo{journal}{\emph{arXiv preprint arXiv:1312.5602}} (\bibinfo{year}{2013}).
\newblock


\bibitem[OpenAI(2022)]%
        {openai2022chatgpt}
\bibfield{author}{\bibinfo{person}{OpenAI}.} \bibinfo{year}{2022}\natexlab{}.
\newblock \bibinfo{title}{{ChatGPT}}.
\newblock
\newblock
\urldef\tempurl%
\url{https://openai.com/blog/chatgpt}
\showURL{%
\tempurl}
\newblock
\shownote{Accessed: [Date you accessed the website]}.


\bibitem[OpenAI(2023)]%
        {openai2023gpt4}
\bibfield{author}{\bibinfo{person}{OpenAI}.} \bibinfo{year}{2023}\natexlab{}.
\newblock \bibinfo{booktitle}{\emph{{GPT-4 Technical Report}}}.
\newblock \bibinfo{type}{{T}echnical {R}eport}.
\newblock


\bibitem[Picasso et~al\mbox{.}(2019)]%
        {2picasso2019technical}
\bibfield{author}{\bibinfo{person}{Andrea Picasso}, \bibinfo{person}{Simone Merello}, \bibinfo{person}{Yukun Ma}, \bibinfo{person}{Luca Oneto}, {and} \bibinfo{person}{Erik Cambria}.} \bibinfo{year}{2019}\natexlab{}.
\newblock \showarticletitle{Technical analysis and sentiment embeddings for market trend prediction}.
\newblock \bibinfo{journal}{\emph{Expert Systems with Applications}}  \bibinfo{volume}{135} (\bibinfo{year}{2019}), \bibinfo{pages}{60--70}.
\newblock


\bibitem[Qian et~al\mbox{.}(2024)]%
        {qian2024chatdev}
\bibfield{author}{\bibinfo{person}{Chen Qian}, \bibinfo{person}{Wei Liu}, \bibinfo{person}{Hongzhang Liu}, \bibinfo{person}{Nuo Chen}, \bibinfo{person}{Yufan Dang}, \bibinfo{person}{Jiahao Li}, \bibinfo{person}{Cheng Yang}, \bibinfo{person}{Weize Chen}, \bibinfo{person}{Yusheng Su}, \bibinfo{person}{Xin Cong}, {et~al\mbox{.}}} \bibinfo{year}{2024}\natexlab{}.
\newblock \showarticletitle{Chatdev: Communicative agents for software development}. In \bibinfo{booktitle}{\emph{Proceedings of the 62nd Annual Meeting of the Association for Computational Linguistics (Volume 1: Long Papers)}}. \bibinfo{pages}{15174--15186}.
\newblock


\bibitem[Qin et~al\mbox{.}(2024)]%
        {2qin2024earnhft}
\bibfield{author}{\bibinfo{person}{Molei Qin}, \bibinfo{person}{Shuo Sun}, \bibinfo{person}{Wentao Zhang}, \bibinfo{person}{Haochong Xia}, \bibinfo{person}{Xinrun Wang}, {and} \bibinfo{person}{Bo An}.} \bibinfo{year}{2024}\natexlab{}.
\newblock \showarticletitle{Earnhft: Efficient hierarchical reinforcement learning for high frequency trading}. In \bibinfo{booktitle}{\emph{Proceedings of the AAAI Conference on Artificial Intelligence}}, Vol.~\bibinfo{volume}{38}. \bibinfo{pages}{14669--14676}.
\newblock


\bibitem[Schulman et~al\mbox{.}(2017)]%
        {schulman2017proximal}
\bibfield{author}{\bibinfo{person}{John Schulman}, \bibinfo{person}{Filip Wolski}, \bibinfo{person}{Prafulla Dhariwal}, \bibinfo{person}{Alec Radford}, {and} \bibinfo{person}{Oleg Klimov}.} \bibinfo{year}{2017}\natexlab{}.
\newblock \showarticletitle{Proximal policy optimization algorithms}.
\newblock \bibinfo{journal}{\emph{arXiv preprint arXiv:1707.06347}} (\bibinfo{year}{2017}).
\newblock


\bibitem[Steinert and Altmann(2023)]%
        {2steinert2023linking}
\bibfield{author}{\bibinfo{person}{Rick Steinert} {and} \bibinfo{person}{Saskia Altmann}.} \bibinfo{year}{2023}\natexlab{}.
\newblock \showarticletitle{Linking microblogging sentiments to stock price movement: An application of GPT-4}.
\newblock \bibinfo{journal}{\emph{arXiv preprint arXiv:2308.16771}} (\bibinfo{year}{2023}).
\newblock


\bibitem[Sujatha~Ravindran and Contreras-Vidal(2023)]%
        {sujatha2023empirical}
\bibfield{author}{\bibinfo{person}{Akshay Sujatha~Ravindran} {and} \bibinfo{person}{Jose Contreras-Vidal}.} \bibinfo{year}{2023}\natexlab{}.
\newblock \showarticletitle{An empirical comparison of deep learning explainability approaches for EEG using simulated ground truth}.
\newblock \bibinfo{journal}{\emph{Scientific Reports}} \bibinfo{volume}{13}, \bibinfo{number}{1} (\bibinfo{year}{2023}), \bibinfo{pages}{17709}.
\newblock


\bibitem[Touvron et~al\mbox{.}(2023)]%
        {touvron2023llama}
\bibfield{author}{\bibinfo{person}{Hugo Touvron}, \bibinfo{person}{Thibaut Lavril}, \bibinfo{person}{Gautier Izacard}, \bibinfo{person}{Xavier Martinet}, \bibinfo{person}{Marie-Anne Lachaux}, \bibinfo{person}{Timoth{\'e}e Lacroix}, \bibinfo{person}{Baptiste Rozi{\`e}re}, \bibinfo{person}{Naman Goyal}, \bibinfo{person}{Eric Hambro}, \bibinfo{person}{Faisal Azhar}, {et~al\mbox{.}}} \bibinfo{year}{2023}\natexlab{}.
\newblock \showarticletitle{Llama: Open and efficient foundation language models}.
\newblock \bibinfo{journal}{\emph{arXiv preprint arXiv:2302.13971}} (\bibinfo{year}{2023}).
\newblock


\bibitem[Wen et~al\mbox{.}(2019)]%
        {2wen2019stock}
\bibfield{author}{\bibinfo{person}{Min Wen}, \bibinfo{person}{Ping Li}, \bibinfo{person}{Lingfei Zhang}, {and} \bibinfo{person}{Yan Chen}.} \bibinfo{year}{2019}\natexlab{}.
\newblock \showarticletitle{Stock market trend prediction using high-order information of time series}.
\newblock \bibinfo{journal}{\emph{Ieee Access}}  \bibinfo{volume}{7} (\bibinfo{year}{2019}), \bibinfo{pages}{28299--28308}.
\newblock


\bibitem[Xie et~al\mbox{.}(2023)]%
        {2xie2023wall}
\bibfield{author}{\bibinfo{person}{Qianqian Xie}, \bibinfo{person}{Weiguang Han}, \bibinfo{person}{Yanzhao Lai}, \bibinfo{person}{Min Peng}, {and} \bibinfo{person}{Jimin Huang}.} \bibinfo{year}{2023}\natexlab{}.
\newblock \showarticletitle{The wall street neophyte: A zero-shot analysis of chatgpt over multimodal stock movement prediction challenges}.
\newblock \bibinfo{journal}{\emph{arXiv preprint arXiv:2304.05351}} (\bibinfo{year}{2023}).
\newblock


\bibitem[Xie et~al\mbox{.}(2024)]%
        {xie2024pixiu}
\bibfield{author}{\bibinfo{person}{Qianqian Xie}, \bibinfo{person}{Weiguang Han}, \bibinfo{person}{Xiao Zhang}, \bibinfo{person}{Yanzhao Lai}, \bibinfo{person}{Min Peng}, \bibinfo{person}{Alejandro Lopez-Lira}, {and} \bibinfo{person}{Jimin Huang}.} \bibinfo{year}{2024}\natexlab{}.
\newblock \showarticletitle{Pixiu: A comprehensive benchmark, instruction dataset and large language model for finance}.
\newblock \bibinfo{journal}{\emph{Advances in Neural Information Processing Systems}}  \bibinfo{volume}{36} (\bibinfo{year}{2024}).
\newblock


\bibitem[Xu et~al\mbox{.}(2023)]%
        {xu2023large}
\bibfield{author}{\bibinfo{person}{Fangzhi Xu}, \bibinfo{person}{Qika Lin}, \bibinfo{person}{Jiawei Han}, \bibinfo{person}{Tianzhe Zhao}, \bibinfo{person}{Jun Liu}, {and} \bibinfo{person}{Erik Cambria}.} \bibinfo{year}{2023}\natexlab{}.
\newblock \showarticletitle{Are large language models really good logical reasoners? a comprehensive evaluation from deductive, inductive and abductive views}.
\newblock \bibinfo{journal}{\emph{arXiv preprint arXiv:2306.09841}} (\bibinfo{year}{2023}).
\newblock


\bibitem[Yang et~al\mbox{.}(2023)]%
        {2yang2023fingpt}
\bibfield{author}{\bibinfo{person}{Hongyang Yang}, \bibinfo{person}{Xiao-Yang Liu}, {and} \bibinfo{person}{Christina~Dan Wang}.} \bibinfo{year}{2023}\natexlab{}.
\newblock \showarticletitle{Fingpt: Open-source financial large language models}.
\newblock \bibinfo{journal}{\emph{arXiv preprint arXiv:2306.06031}} (\bibinfo{year}{2023}).
\newblock


\bibitem[Yu et~al\mbox{.}(2023)]%
        {2yu2023temporal}
\bibfield{author}{\bibinfo{person}{Xinli Yu}, \bibinfo{person}{Zheng Chen}, \bibinfo{person}{Yuan Ling}, \bibinfo{person}{Shujing Dong}, \bibinfo{person}{Zongyi Liu}, {and} \bibinfo{person}{Yanbin Lu}.} \bibinfo{year}{2023}\natexlab{}.
\newblock \showarticletitle{Temporal Data Meets LLM--Explainable Financial Time Series Forecasting}.
\newblock \bibinfo{journal}{\emph{arXiv preprint arXiv:2306.11025}} (\bibinfo{year}{2023}).
\newblock


\bibitem[Yu et~al\mbox{.}(2024)]%
        {2yu2024finmem}
\bibfield{author}{\bibinfo{person}{Yangyang Yu}, \bibinfo{person}{Haohang Li}, \bibinfo{person}{Zhi Chen}, \bibinfo{person}{Yuechen Jiang}, \bibinfo{person}{Yang Li}, \bibinfo{person}{Denghui Zhang}, \bibinfo{person}{Rong Liu}, \bibinfo{person}{Jordan~W Suchow}, {and} \bibinfo{person}{Khaldoun Khashanah}.} \bibinfo{year}{2024}\natexlab{}.
\newblock \showarticletitle{FinMem: A performance-enhanced LLM trading agent with layered memory and character design}. In \bibinfo{booktitle}{\emph{Proceedings of the AAAI Symposium Series}}, Vol.~\bibinfo{volume}{3}. \bibinfo{pages}{595--597}.
\newblock


\bibitem[Zhang et~al\mbox{.}(2023)]%
        {2zhang2023enhancing}
\bibfield{author}{\bibinfo{person}{Boyu Zhang}, \bibinfo{person}{Hongyang Yang}, \bibinfo{person}{Tianyu Zhou}, \bibinfo{person}{Muhammad Ali~Babar}, {and} \bibinfo{person}{Xiao-Yang Liu}.} \bibinfo{year}{2023}\natexlab{}.
\newblock \showarticletitle{Enhancing financial sentiment analysis via retrieval augmented large language models}. In \bibinfo{booktitle}{\emph{Proceedings of the fourth ACM international conference on AI in finance}}. \bibinfo{pages}{349--356}.
\newblock


\bibitem[Zhang et~al\mbox{.}(2022a)]%
        {2zhang2022decision}
\bibfield{author}{\bibinfo{person}{Cheng Zhang}, \bibinfo{person}{Nilam~NA Sjarif}, {and} \bibinfo{person}{Roslina~B Ibrahim}.} \bibinfo{year}{2022}\natexlab{a}.
\newblock \showarticletitle{Decision fusion for stock market prediction: a systematic review}.
\newblock \bibinfo{journal}{\emph{IEEE Access}}  \bibinfo{volume}{10} (\bibinfo{year}{2022}), \bibinfo{pages}{81364--81379}.
\newblock


\bibitem[Zhang et~al\mbox{.}(2024a)]%
        {zhang2024agentpro}
\bibfield{author}{\bibinfo{person}{Wenqi Zhang}, \bibinfo{person}{Ke Tang}, \bibinfo{person}{Hai Wu}, \bibinfo{person}{Mengna Wang}, \bibinfo{person}{Yongliang Shen}, \bibinfo{person}{Guiyang Hou}, \bibinfo{person}{Zeqi Tan}, \bibinfo{person}{Peng Li}, \bibinfo{person}{Yueting Zhuang}, {and} \bibinfo{person}{Weiming Lu}.} \bibinfo{year}{2024}\natexlab{a}.
\newblock \showarticletitle{Agent-pro: Learning to evolve via policy-level reflection and optimization}.
\newblock \bibinfo{journal}{\emph{arXiv preprint arXiv:2402.17574}} (\bibinfo{year}{2024}).
\newblock


\bibitem[Zhang et~al\mbox{.}(2024b)]%
        {zhang2024finagent}
\bibfield{author}{\bibinfo{person}{Wentao Zhang}, \bibinfo{person}{Lingxuan Zhao}, \bibinfo{person}{Haochong Xia}, \bibinfo{person}{Shuo Sun}, \bibinfo{person}{Jiaze Sun}, \bibinfo{person}{Molei Qin}, \bibinfo{person}{Xinyi Li}, \bibinfo{person}{Yuqing Zhao}, \bibinfo{person}{Yilei Zhao}, \bibinfo{person}{Xinyu Cai}, {et~al\mbox{.}}} \bibinfo{year}{2024}\natexlab{b}.
\newblock \showarticletitle{FinAgent: A Multimodal Foundation Agent for Financial Trading: Tool-Augmented, Diversified, and Generalist}.
\newblock \bibinfo{journal}{\emph{arXiv preprint arXiv:2402.18485}} (\bibinfo{year}{2024}).
\newblock


\bibitem[Zhang et~al\mbox{.}(2022b)]%
        {zhang2022automatic}
\bibfield{author}{\bibinfo{person}{Zhuosheng Zhang}, \bibinfo{person}{Aston Zhang}, \bibinfo{person}{Mu Li}, {and} \bibinfo{person}{Alex Smola}.} \bibinfo{year}{2022}\natexlab{b}.
\newblock \showarticletitle{Automatic chain of thought prompting in large language models}.
\newblock \bibinfo{journal}{\emph{arXiv preprint arXiv:2210.03493}} (\bibinfo{year}{2022}).
\newblock


\bibitem[Zhu et~al\mbox{.}(2022)]%
        {zhu2022solving}
\bibfield{author}{\bibinfo{person}{Xinyu Zhu}, \bibinfo{person}{Junjie Wang}, \bibinfo{person}{Lin Zhang}, \bibinfo{person}{Yuxiang Zhang}, \bibinfo{person}{Yongfeng Huang}, \bibinfo{person}{Ruyi Gan}, \bibinfo{person}{Jiaxing Zhang}, {and} \bibinfo{person}{Yujiu Yang}.} \bibinfo{year}{2022}\natexlab{}.
\newblock \showarticletitle{Solving math word problems via cooperative reasoning induced language models}.
\newblock \bibinfo{journal}{\emph{arXiv preprint arXiv:2210.16257}} (\bibinfo{year}{2022}).
\newblock


\end{thebibliography}
\appendix
\section{Appendix}
\label{app}
\begin{table*}[t]
\centering
\begin{tabular}{|p{\textwidth}|}
\hline
\texttt{\textbf{News Summarizer Agent Prompt}} \\
\hline
\texttt{Analyze the following financial news about \{ticker\} and provide a concise summary focused on potential trading implications:} \\[0.5em]
\texttt{1. Summarize the key points in 5-6 sentences, highlighting any information that could impact \{ticker\}'s stock price in the short term.} \\[0.2em]
\texttt{2. Identify any positive and negative factors mentioned in the news.} \\[0.2em]
\texttt{3. Based on this news, would you expect the overall sentiment towards \{ticker\} stock to be bullish, bearish, or neutral? Briefly explain why.} \\[0.5em]
\texttt{News data: \{news\_data\}} \\[0.5em]
\texttt{Please provide your analysis in a clear, concise manner.} \\
\hline
\texttt{\textbf{Chart Analyst Agent Prompt}} \\
\hline
\texttt{Analyze this \{ticker\} candlestick chart focusing on three trading strategies:} \\[0.3em]
\texttt{1. MACD Crossover Strategy} \\
\texttt{2. KDJ with RSI Filter Strategy} \\[0.3em]
\texttt{For each strategy, provide:} \\
\texttt{1. One key market trend observation} \\
\texttt{2. One potential trading signal} \\[0.3em]
\texttt{Format your response as:} \\
\texttt{Strategy 1: [Trend] | [Signal]} \\
\texttt{Strategy 2: [Trend] | [Signal]} \\[0.3em]
\texttt{Be concise. Limit each observation and signal to 10 words or less.} \\
\hline
\texttt{\textbf{Reflection Agent Prompt (Short/Medium-Term)}} \\
\hline
\texttt{Analyze this \{len\_short/medium\_term\_data\}-day \{ticker\} stock trading data:} \\
\texttt{\{json\_data\}} \\[0.3em]
\texttt{Focus on:} \\
\texttt{1. Recent recommendations and their outcomes (reward)} \\
\texttt{2. Key factors influencing decisions} \\
\texttt{3. Cumulative return trend} \\[0.3em]
\texttt{Provide three insights:} \\
\texttt{1. Decision effectiveness (based on recommendations and rewards)} \\
\texttt{2. Most impactful key factors} \\
\texttt{3. Short/medium-term return trend} \\[0.3em]
\texttt{Format: [Effectiveness] | [Key Factors] | [Return Trend]} \\
\texttt{Keep each insight under 15 words.} \\
\hline
\texttt{\textbf{Trading Signal Chart Reflection Prompt}} \\
\hline
\texttt{Analyze this \{ticker\} trading chart showing closing prices and previous trading signals:} \\[0.3em]
\texttt{- The chart displays closing prices over time} \\
\texttt{- Green markers indicate previous BUY decisions} \\
\texttt{- Red markers indicate previous SELL decisions} \\
\texttt{- Absence of markers indicates HOLD decisions} \\[0.3em]
\texttt{Task:} \\
\texttt{1. Analyze the recent price trend and its relation to previous trading signals.} \\
\texttt{2. Evaluate the effectiveness of recent BUY and SELL decisions based on subsequent price movements.} \\
\texttt{3. Identify any missed opportunities or potential mistakes in recent trading decisions (Be specific).} \\[0.3em]
\texttt{Provide your analysis in a concise, bullet-point format.} \\
\hline
\end{tabular}
\caption{Prompt Message Utilized in Agents}
\label{restPrompt}
\end{table*}

\begin{table*}[t]
\centering
\begin{tabular}{|p{\textwidth}|}
\hline
\texttt{\textbf{Prediction Agent Prompt}} \\
\hline
\texttt{As an advanced trading strategy agent for \{ticker\} stock, analyze the following data to formulate an opportunistic trading decision:} \\[0.2em]
\texttt{Current Date: \{state['date'][-1]\}} \\
\texttt{Current Portfolio:} \\
\texttt{- Shares: \{current\_shares\}} \\
\texttt{- Share Price: \$\{current\_price:.2f\}} \\
\texttt{- Average Purchase Price: \$\{avg\_purchase\_price:.2f\}} \\
\texttt{- Total Value: \$\{total\_value:.2f\}} \\
\texttt{- Cash Reserve: \$\{cash\_reserve:.2f\}} \\
\texttt{- Cash Percentage: \{cash\_percentage:.2f\}\%} \\
\texttt{- Unrealized P/L: \$\{(current\_price - avg\_purchase\_price) * current\_shares:.2f\}} \\
\texttt{- Unrealized Profit Percentage: \{unrealized\_profit\_percentage:.2f\}\%} \\[0.2em]
\texttt{Technical Analysis: \{state['chart\_analysis'][-1]\}} \\[0.2em]
\texttt{News Summary: \{state['news\_summary'][-1]\}} \\[0.2em]
\texttt{Reflections:} \\
\texttt{- Short-term: \{state['reflection\_insights']['short\_term']\}} \\
\texttt{- Medium-term: \{state['reflection\_insights']['medium\_term']\}} \\[0.2em]

\texttt{Effectiveness of past trading decisions: \{state['market\_intelligence']\}} \\[0.2em]
\texttt{Historical Trading Data (Last \{len(historical\_data)\} days): \{json\_data\}} \\[0.2em]
\texttt{Based on this data, provide an opportunistic trading strategy. Consider:} \\
\texttt{1. Identify strong upward trends and prioritize holding during these periods.} \\
\texttt{2. Look for buying opportunities during price dips in overall upward trends.} \\
\texttt{3. Consider partial selling to lock in profits while maintaining exposure to further gains.} \\
\texttt{4. Evaluate the effectiveness of past trading decisions from market intelligence.} \\
\texttt{5. Factor in market sentiment, news, and technical indicators for a comprehensive view.} \\
\texttt{6. Balance short-term opportunities with long-term growth potential.} \\[0.2em]
\texttt{Rules:} \\
\texttt{- Maintain at least 10\% of the portfolio in cash for opportunistic buying.} \\
\texttt{- Recommend BUY if:} \\
\texttt{  a) Strong upward trend and sufficient cash (>10\% of portfolio)} \\
\texttt{  b) Price dip in overall upward trend and cash reserve above 20\%} \\
\texttt{- Recommend SELL if:} \\
\texttt{  a) Signs of significant trend reversal, OR} \\
\texttt{  b) Exceptional gains (>5\% from avg purchase price) for partial sell, OR} \\
\texttt{  c) Consistent gains for 3 or more consecutive days} \\
\texttt{- Recommend HOLD if:} \\
\texttt{  a) Upward trend continuing without significant reversal signs} \\
\texttt{  b) Market uncertainty and current positions are profitable} \\
\texttt{- For BUY: 5-10\% in strong upward trends, 1-5\% for dip opportunities} \\
\texttt{- For SELL: Consider partial sells (3-5\%) to lock in profits} \\
\texttt{- Be more aggressive with SELL during strong upward trends} \\[0.2em]
\texttt{Provide your trading strategy in the following format:} \\
\texttt{Recommendation: [BUY/SELL/HOLD]} \\
\texttt{Position Size: [1-10] (0 if HOLD)} \\
\texttt{Explanation: [Detailed rationale for the decision]} \\
\texttt{Ensure your explanation is detailed and covers all aspects of your analysis and decision-making process, with a focus on capturing opportunistic gains.} \\
\hline
\end{tabular}
\caption{Prediction Agent Prompt}
\label{predictiom}
\end{table*}

\end{document}